\documentstyle [pre,aps,epsfig]{revtex}
\newenvironment{mytitle}{\begin{center} \large \bf }{\\ [.1in]\end{center}}
\newenvironment{myauthor}{\begin{center} }{\\ [.1in]\end{center}} 
\newenvironment{myinstit}{\begin{center} \it}{\end{center}}
\begin{document}
\thispagestyle{empty}
\begin{mytitle}
 Quantum-classical correspondence for local density of states and eigenfunctions of
a chaotic periodic billiard
\end{mytitle}

\begin{myauthor}
G. A. Luna-Acosta*,  J. A. M\'endez-Berm\'udez, and F. M. Izrailev
\end{myauthor}
\begin{myinstit}
Instituto de F\'{\i}sica,
Universidad Aut\'onoma de Puebla, Puebla, Pue. 72570,
Apartado Postal J - 48,
M\'exico.
\end{myinstit}
\vspace*{.5cm}
\hrule
\vspace*{-.8cm}
\begin{abstract} 
\flushleft {\bf Abstract}

Classical-quantum correspondence for conservative chaotic Hamiltonians is investigated in 
terms of the structure of the eigenfunctions and the local density of states, using as a model a 2D rippled billiard in the regime of global chaos. The influence of the observed localized and sparsed states in the quantum-classical correspondence is discussed.\\

PACS: 5.45.+b; 03.65.-w; 03.20.\\

Keywords: Quantum-classical correspondence, chaotic billiards.

* Corresponding author. E-mail:  gluna@sirio.ifuap.buap.mx
\end{abstract}
\vspace*{.3 cm}
\hrule

\vspace*{.5cm}

The subject of this Letter concerns  the  quantum-classical correspondence of autonomous Hamiltonian
systems with classical chaotic dynamics.  As a paradigm for this class of  systems
we  consider the motion of particles in  the 2D rippled billiard depicted in Fig. 1. It is known [1-3] that the dynamics of a classical particle in this billiard undergoes the generic transition to chaos (regular-mixed-global) as the amplitude $a$ is increased ($d$ is fixed). Thus the results that can be obtained by studying this particular model are applicable to a wide
class of systems. A modern physical realization of this rippled billiard of {\it finite} length is a mesoscopic electron wave guide.
In [1] its {\it classical}  transport properties, such as resistivity, were related to its dynamical properties, yielding a 
transport signature of chaos. The analysis of the quantum motion in the {\it infinitely} long rippled channel (the periodic rippled channel) is useful for the understanding of universal features of  electronic band structures of real crystals [4], propagation in periodic  waveguides [5], quantum wires [6-8]
and films [9]. Signatures of chaotic diffusion in the  band spectra of a  similar periodic billiard have been 
investigated  in Ref. [10].
 In [2] the energy band structure of the quantum version of the periodic rippled channel
 was calculated and certain aspects of the quantum-classical correspondence were investigated in terms
of the Husimi distributions. \\

\begin{figure}[htb]
\begin{center}
\epsfig{file=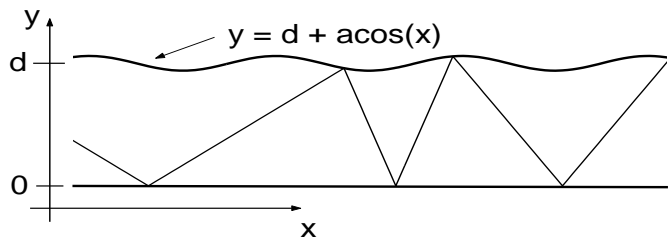,width=3.5in,height=1.3in}
\caption{Geometry of the rippled channel.}
%\label{weaal0}
\end{center}
\end{figure}

 \indent The standard signature of chaos in quantum mechanics [11] based on the
random matrix conjecture [12] showed a qualitative agreement for the  energy level spacing statistics
of the periodic rippled channel [13]. However, a more detailed study [14] revealed important deviations.
Motivated by this discrepancy, which may be found in other chaotic systems, here we use a novel
approach that has been 
suggested in Ref. [15] to analyze the quantum-classical correspondence of complex systems. Such an approach relies  
on the study of the structure of the local density of states (LDOS),
and the shape  of the eigenfunctions (SEF) in the basis of non-interacting particles (quasi-particles) [16]. 
It has been  confirmed that the LDOS and SEF have well defined classical counterparts
in the chaotic regime in dynamical systems such as that of  two interacting spin particles [17] and the three
 orbital schematic 
shell model [18], see also very recent works in [19].
Below we extend these studies to chaotic billiards by treating the degrees of freedom as 
independent particles and incorporating the effects of the boundaries into the Hamiltonian operator. We show 
how to construct the classical counterparts of LDOS and SEF for chaotic billiards. We shall demonstrate that these
quantities determine the global shape of the quantum LDOS and SEF, and that the phenomenon of  quantum localization 
within the
energy shell manifests itself as strong fluctuations about the global shape. We remark here that this localization occurs
not in configuration space but in energy (or momentum) space (for chaotic billiards this kind of localization 
was first investigated in Ref. [20]).

We limit our studies  here to the case of global classical chaos which occurs in wide ($d>2\pi$) channels [21].
The control parameter that determines the degree of chaoticity, obtained by the standard
linearization of its mapping, is given by
$K=\frac{2 da}{\pi}$. In agreement with Chirikov's overlapping criteria our numerical experiments show global
chaos for $K \stackrel{>}{\sim}1$. We have compared the phase space dynamics for various combinations of $a$ and
$d$ yielding  the same value of $K>1$, and found their Poincare maps to be indistinguishable from each other.
In what follows we take  ($a/2\pi, d/2\pi)= (.06,1)$, i.e.,  $K\approx 1.5$.\\ 

The quantum mechanical description involves the solution of the Schr\"odinger equation $\hat H=\frac{\hbar^2}{2}(\hat p_x^2 +\hat p_y^2)$ subject to the boundary conditions $\psi(x,y)=0$ at $y=0$ and $y=d+a \cos x$. It is both  convenient and illuminating to go to the new coordinates $u=x$, and $v=\frac{yd}{d + a cos( x)}$, for in these coordinates the boundary conditions become simpler: $\Psi(u,v) = 0$ at $v = 0, d$. On the other hand, the Hamiltonian acquires a much more complicated form,

\begin{equation}
\hat H= \frac{\hbar^2}{2} g^{-1/4}\hat P_\alpha g^{\alpha \beta} g^{1/2} \hat P_\beta g^{-1/4}, \ \ \alpha, \beta=u,v,
\end{equation}

\noindent which is simply the kinetic energy expressed in covariant form [22]. The momentum
is now given by $\hat P_\alpha=-i \hbar[\partial_\alpha + \frac{1}{4} \partial_\alpha ln(g)]=
-i\hbar g^{-1/4}\partial_\alpha g^{1/4}$, where  $g^{\alpha \beta}$ is the metric tensor, and
$g= Det(g_{\alpha \beta})=[1 + \epsilon \cos u]^2$, $\epsilon\equiv a/d$, is the metric (see [2] for details). Thus, we have formally transformed the model of one-particle motion in the rippled billiard to that of two interacting particles (identified with the two degrees of freedom). In this way, the complexity of the boundary in the original model is incorporated into the interaction potential.

Since the Hamiltonian is periodic in $u$, the energy eigenstates satisfy Bloch's theorem: $\Psi_E(u,v)=
 \exp (iku)\Phi_k(u,v)$ with $ \Phi_k(u+2\pi,v)=\Phi_k(u,v)$ and Bloch wave vector $k=k(E)$. For an infinite channel, $k$ takes
a continuous range of values and we choose the first Brillouin zone to lie in the interval $[-\frac{1}{2}\leq k\leq\frac{1}{2}]$.
Further, we can expand $ \Phi_k(u,v)$  in a double Fourier series and write the $\alpha^{th}$ eigenstate of energy
$E^{\alpha}(k)$ as 
\begin{equation}
\Psi^{\alpha}(u,v;k)= \sum_{m=1}^{\infty}\sum_{n=-\infty}^{\infty}C_{mn}^{\alpha}(k) \phi_{mn}^k(u,v)
\end{equation}
where
\begin{equation}
\phi_{mn}^k(u,v)=<u,v \mid m,n>_k=\pi^{-1/2}g^{-1/4}sin(m \pi v) e^{i(k+n)u}.
\end{equation}

The factor  $\pi^{-1/2} g^{-1/4}$ comes from the orthonormality condition in the curvilinear coordinates $(u,v)$.
For future reference we shall call $m$ the mode (or channel) number and $n$ the Brillouin Zone number.
Note that $\phi_{mn}^k$  are the eigenstates  of the unperturbed momenta squared (and, therefore, of the unperturbed Hamiltonian):
 $\hat P_v^2 \phi_{mn}^k=
\hbar^2(m\pi)^2\phi_{mn}^k$, $\hat P_u^2 \phi_{mn}^k  =  \hbar^2(k+n)^2\phi_{mn}^k$. 
 The problem requires the diagonalization of
the matrix $H_{l,l'}(k)=<l\mid\hat H\mid l'>_k$, where $ \mid l>_k=\mid m,n>_k$. Note that  for each pair $(m,n)$ we  
associate a number $l$. Even though the energy spectra is independent of the choice of the function $l(m,n)$,
 the structure of the matrix clearly depends on it.  For our purposes, it is essential (as it will become clear below) that
the unperturbed Hamiltonian basis be ordered in increasing energy:  $E_{l+1}^0(k)>E_l^0(k)$, where $H^0 \phi_l^k=E_l^0(k) \phi_l^k$,
$E_l^0(k)=\frac{\hbar^2}{2}((n+k)^2 + (m\pi)^2)$.  This defines the choice of assignment $(m,n)\Rightarrow l$.
Once the matrix $H_{l,l'}(k)$ has been diagonalized, its (exact) eigenstates $\Psi^{\alpha}(k)=\sum C_{\ l}^{\alpha}(k) \phi^k_l$
are also re-ordered in energy $(E^{\alpha +1}\geq E^{\alpha})$.  We adopt  the convention that the Greek
superindex (Latin subindex) denotes the exact (unperturbed) state. The amplitudes $C^{\alpha}_{\ l}(k)$ form the 
state vector matrix;
its elements along the row $\alpha$ are the components of the $ \alpha^{th}$  state in the representation of the
unperturbed re-ordered basis, and the elements along the  column $l$ give the unperturbed state $l$ expanded in
the re-ordered perturbed basis.
We shall study the structure of both, rows and columns, the latter being essential in the construction of the local
density of states (LDOS) of the system.
 
 In the previous work [13] we studied the level spacing statistics of the rippled channel in connection with the 
Random Matrix Theory conjecture. Using the Bloch momentum $k$ as an external parameter 
we found that the level spacing statistics is the same for all values of $k$ (except $k\approx 0$) {\it within} the Brillouin zone.
Thus, here we arbitrarily take the value $k=0.1$ to explore the structure of the exact eigenstates
and from now on we drop, for economy of notation,  the label $k$.

\begin{figure}[htb]
\begin{center}
\epsfig{file=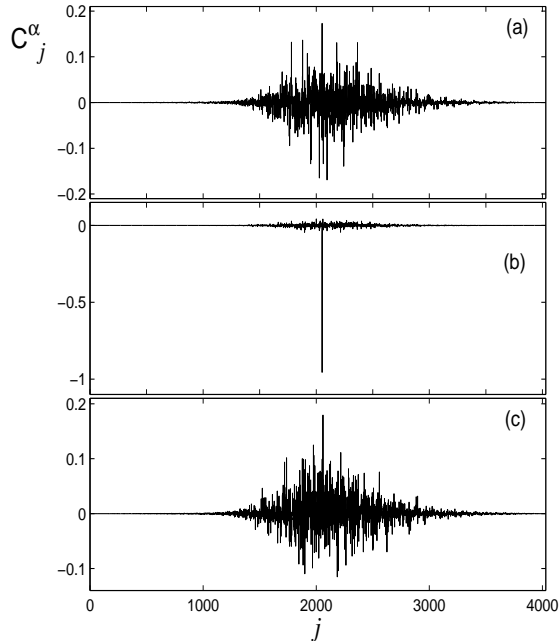,width=2.9in,height=3.4in}
\caption{Typical high energy eigenfunctions for a) $\alpha$ = 2078, b) $\alpha $ = 2079 and c) $\alpha$ = 2080. The localization measures
are: a) $l_H$ = 653.5, $l_{ipr}$ = 411.5; b) $l_H$ = 4.5, $l_{ipr}$ = 3.6; and c) $l_H$ = 747.8, $l_{ipr}$ = 571.7. The amplitude and width
of the channel are $a/2\pi$ = .06 and $d/2\pi$ = 1, respectively. The value of the Bloch wave vector is $k = 0.1$.}
%\label{weaal0}
\end{center}
\end{figure}

Fig. 2 shows three consecutive high-energy eigenstates ($C^{\alpha}_{\ j}$, $\alpha$=2078, 2079, and 2080) in the re-ordered unperturbed basis. This sequence illustrates the kind of eigenstates that can occur typically; in a given sequence for this range of values of $\alpha$ roughly 15 percent are states such as the one shown in Fig. 2b, and about 60 percent are like those of Figs. 2a and 2c. The rest are intermediate between these two types: ``sparse" states characterized by a few dominant peaks in a background of smaller components. Eigenstates can be characterized by  various localization measures, such as the ``entropy localization length" $l_{\cal H}$ and the ``inverse participation length" $l_{ipr}$ (see, e.g., Ref. [17]) defined by
\begin{equation}
l_{\cal H}= 2.08 \exp(-{\cal H}), \         \   l_{ipr}= 3/{\cal P},
\end{equation}
where
\begin{equation}
{\cal H}=\sum^N_{\alpha=1}\mid \psi^\alpha(E)\mid^2 ln \mid \psi^\alpha(E)\mid^2,  \   \ and \ \ {\cal P}= \sum^N_{\alpha=1} \mid \psi^\alpha(E) \mid^4.
\end{equation}

 These two measures are found to be two orders of
magnitude smaller for the eigenstate 2079 than for the eigenstates 2078 and 2080 shown in Fig. 2.
 Extensive numerical
results [23] show that these lengths fluctuate wildly as a function of the number of the eigenstate $\alpha$,
or, equivalently, as a function of its corresponding energy $E^{\alpha}$.
As for extremely localized states such as that of Fig. 2b, they 
are expected to play an important role in transport since they are practically the same as the unpertubed (regular) states
of the model. In the energy spectrum, they are located at the edges  of the cells defined by different values of the
Brillouin Zone number $n$. Moreover, these states have the largest group velocity $\partial E/ \partial k$; see [23]
for details.

%\section{Classical EFS and LDOS}
Our main interest here is to relate the structure of the matrix $C^{\alpha}_{ \ j}$ to properties of the corresponding
classical system.
Since $C^{\alpha}_{\ j} = < \Psi^{\alpha}\mid \phi_j>$ as a function of $j$ (or equivalently, $E^0_ {\ j}$)
is the {\it projection} of the exact state onto the states of the unperturbed system,
a classical counterpart of $ \mid C^{\alpha}_{ \ j} \mid^2$ as a function of $j$ can be obtained [15] by {\it projecting}
 the (chaotic) trajectories generated by the 
Hamiltonian $H$ onto the unperturbed Hamiltonian $H^0$, where $H=H^0 +V$ and $V$ is the perturbation. 
Specifically, this can be done by  substituting the orbit $\Phi(t)\equiv (x(t), y(t), p_x(t), p_y(t))$ of a typical initial condition
generated by $H$ 
with energy $E$ onto $H^0$. Then the energy of the unperturbed Hamiltonian varies in time, 
$E^0(t)$, fluctuating within a range $\Delta E$ which defines the width of
 the so-called  ``energy shell." After a sufficiently long
time a distribution of energies $W_{cl}(E^0/E)$ can then be obtained from $E^0(t)$ for a single (chaotic) orbit. Alternatively, 
$W_{cl}(E^0/E)$ can be calculated for shorter times but taking several orbits. The equivalence between these two
ways is expected if motion generated by the full Hamiltonian is globally chaotic.

The above discussion outlines the general procedure that claims that $W_{cl}(E^0/E)$ is the classical counterpart
of $ \mid C^{\alpha}_{\ j} \mid^2$ as a function of $E^0_j$.  
It has been applied  [15-19] in a straight forward manner in systems
where $H^0$ and $V$ can readily be identified. 
For chaotic billiards (with Neumann or Dirichlet boundary conditions) the Hamiltonian of both unperturbed and perturbed 
systems is simply the kinetic energy and the non-integrability comes from the boundary conditions. 
In order to incorporate the perturbation into the Hamiltonian operator, it is necessary to perform an appropriate coordinate transformation in the same way as it was done above for the quantum model. To accomplish this, we take the quantum Hamiltonian, Eq. (1), and
commute all the operators, transforming them to c-numbers,
\begin{equation}
H= \frac{1}{2} g^{\alpha \beta} P_{\alpha}P_ {\beta}, 
\end{equation}
subject to the ``flat" boundary conditions in $(u,v)$ coordinates.
Expanding this expression, we obtain

\begin{equation}
H=\frac{1}{2}\left[ P_u^2- \frac{2 \epsilon v \xi _u}{1+\epsilon \xi} P_u P_v + \frac{1+(\epsilon v \xi_u)^2}{(1+\epsilon \xi)^2}P_v^2 \right],
\end{equation}

where $\xi\equiv cosu, \ \xi_u\equiv d\xi/du$, and $\epsilon\equiv a/d$. The canonical transformation from the old to new variables is given by $u=x$, $v=yd/(d+a \xi)$, $P_u=P_x +  \epsilon y \xi_x P_y/(1 + \epsilon \xi)$, and $P_v=(1+ \epsilon \xi)P_y$.
We now separate the Hamiltonian $H=H^0 +V$ analogously to the quantum model: $H^0\equiv  \frac{1}{2} \left[ P_u^2 +P_v^2  \right]; \,\,\,\,  V \equiv H-H^0$.\\

\indent To compare the classical distribution $W_{cl}(E^0/E)$  with the quantum one 
$\mid C^\alpha_{\ j} \mid^2$ (as a function of $E^0_{\ j}$) we take the energy of the classical Hamiltonian to be equal
to the energy of the state $\mid \Psi^\alpha>$ under consideration. Since the {\it classical} phase space dynamics is independent
of the energy of the particle (the kinetic energy can be rescaled) we arbitrarily take it to be equal to one and introduce
an effective Planck's constant $\hbar_{eff}$.\\
 In Fig. 3 we show the classical and quantum distributions (shapes) of four eigenstates.
The areas under the classical and quantum curves have been normalized to 1.
These states were chosen from four different energy regions (different values of $\hbar_{eff}$) and they represent typical states in each region of about one hundred eigenstates. These are typical in the sense that most states in each
interval are characterized by  typical localization lengths in that range. For example in the region of $700<\alpha<800$, the
length $l_{\cal H}$ oscillates around $l_{\cal H}=200$ with only 8 eigenstates with $l_{\cal H}<50$ and
the maximum of $l_{\cal H}$ is about 340. That is, the states of Fig. 3 are neither extremely delocalized nor extremely localized.

  A simple inspection of Fig. 3 indicates that
for low energies few unperturbed components participate in the construction  of the exact eigenstate and as the energy
increases, the participation of more and more components increases. Based on the Weyl
formula and the condition that the de Broglie wavelength be equal to the amplitude of the ripple, one obtains an estimate
for the number of levels defining the perturbative border $N_p$. For the geometrical parameters of the channel
 ($a/2\pi=.06, d/2\pi=1$) we have $N_p\approx 139$. Thus, for low energies, as exemplified by Fig. 3a,  there is 
hardly any resemblance between the classical and quantum distributions, since at such  
energies the system is still in the perturbative regime and far from the semiclassical domain.
It is also clear that as
 the energy increases the global
shape of the quantum distribution is closer to the classical distribution.
It is remarkable that the range of energy components that participate in the construction of the perturbed eigenstate can 
be deduced from the classical distribution. 
This range of energies defines the {\it energy shell} of the eigenstate under consideration. Note that the slight assymmetry observed in the range of the quantum distribution with respect to $E^0=1$ is the same as in the classical distribution and can be explained analytically, see details in Ref. [23].

\begin{figure}[htb]
\begin{center}
\epsfig{file=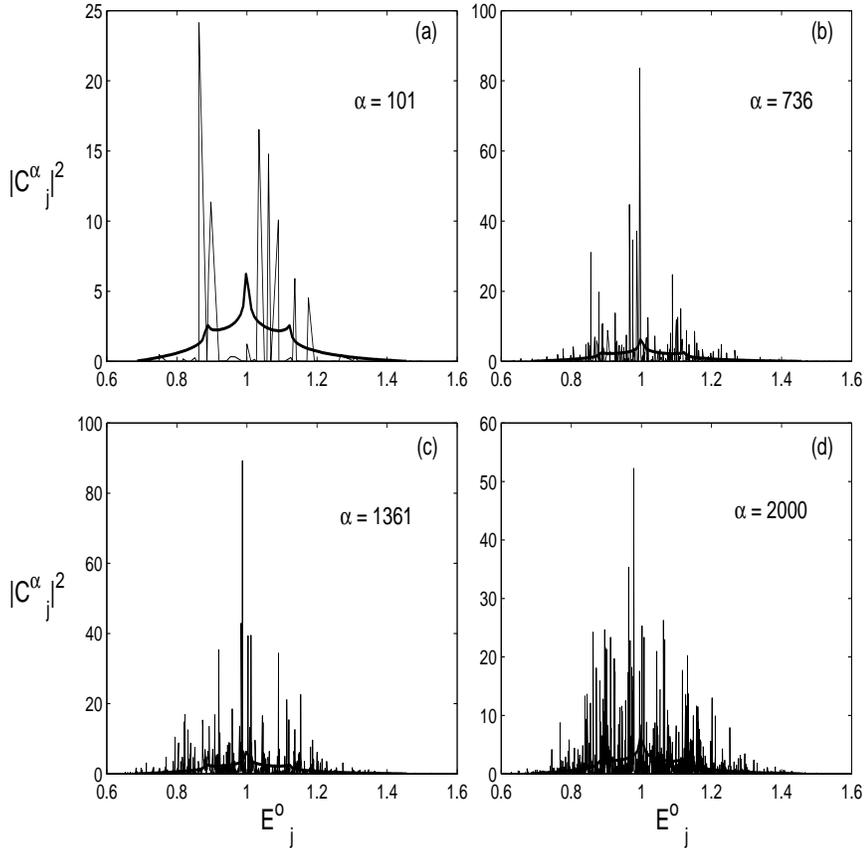,width=4.5in,height=4.5in}
\caption{Individual eigenfunctions $ \mid C^\alpha_{\ j}\mid^2 $ in the energy representation and the classical distribution $ W_{cl}(E^0/E) $ (thick curve) as a function of the energy $ E^0_{\ j} $ for $\alpha$= a) 101; b) 736; c)1361; and c)2000. The entropy localization length $l_{\cal H}$ is a) 18.7; b) 162.7; c) 422.3; and d) 707.2. The effective Planck's constant $\hbar_{eff}$ is a) 0.24, b) 0.09, c) 0.07, and d) 0.05. Same values of amplitude $a$, width $d$, and $k$ as in Fig. 2. Note the different vertical scales due to normalization conditions.}
%\label{weaal0}
\end{center}
\end{figure}

For states with largest values of $l_{ipr}$ or $l_{\cal H}$, our numerical experiments show a much better quantum-classical correspondence.
This occurs because as more components are needed to form the
perturbed state, then the number of  components with amplitudes above the classical shape decreases.
Even so, strong fluctuations of the quantum distribution with respect to the classical shape do not dissappear completely
even for the case of extremely delocalized (within the energy shell) states, see Fig. 3d.\\

 We now try to smooth these fluctuations
 by calculating the averaged distribution, also known as the average shape of the eigenfunction (SEF)
\begin{equation}
W(E^0_{\  j}/E^\alpha) =  \sum_{\alpha'} \mid C^{\alpha'}_{\  j}\mid^2 \delta(E-E^{\alpha'}),
\end{equation}

where the sum is taken over a number of eigenstates in the neighborhood of the exact state $\mid \Psi^\alpha>$.
%Since the center and the width of the energy shell may strongly fluctuate, due to quantum effects, the sum in the equation above is carried out by re-centering each eigenstate to coincide with the
%centroid of the state $\mid \Psi^\alpha>$.

Another quantum quantity that shall be compared with its classical counterpart is the so called ``local density of states"  (LDOS)
or ``strength
function", widely used in solid state and nuclear physics. It is defined as
\begin{equation}
\omega(E^\alpha/E^0_{\  j})=\sum_{j'} \mid C^{\alpha}_{\ j'} \mid ^2 \delta (E^0 -E_{j'}),
\end{equation}
where the summation is done over a number of perturbed states in the neighborhood of the unperturbed state 
$\mid j>$.
This quantity gives information about the evolution of the system. In particular, the spreading width of this
function determines the energy range associated with its ``decay'' into other states due to the interaction.
% It is convenient
% to  re-center the unperturbed states in the same way as for the EF, where the centroid
% now refers to the centroid of the unperturbed state $\mid \phi_j>$.
The classical counterpart of the LDOS is constructed in much the same way as for the ``classical eigenfunction" 
distribution $W_{cl}(E^0/E)$ except now we project the trajectories of the unperturbed Hamiltonian $H^0$ onto the
exact Hamiltonian $H$ [24].

\begin{figure}[htb]
\begin{center}
\epsfig{file=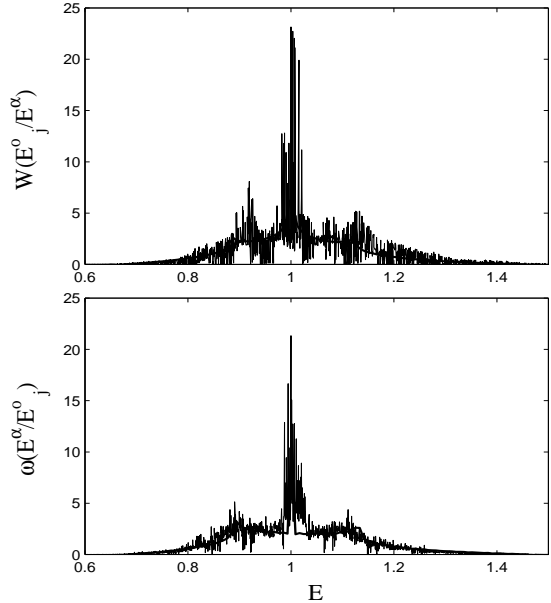,width=2.9in,height=3.2in}
\caption{a) Shape of eigenfuctions (SEF) in the energy representation $W(E^0_{\  j}/E^\alpha)$ and its classical counterpart $W_{cl}(E^0/E)$ (thick curve); b) LDOS $\omega(E^\alpha/E^0_{\  j})$ and its classical counterpart $\omega_{cl}(E/E^0)$ (thick curve). Same values of amplitude $a$, width $d$, and $k$ as in Fig. 2. For the SEF the average is taken over the interval $1870<\alpha<1950$ and for the LDOS the average is over the interval $1900< j <1990$.}
%\label{weaal0}
\end{center}
\end{figure}

Figures 4a and 4b show both distributions, the average SEF and the LDOS, together
with their classical counterparts. Even though the agreement  between classical and averaged quantum eigenfunction shapes 
has improved compared to the individual shapes of Fig. 3, the quantum
fluctuations are still relatively large. Moreover, the averaging procedure reveals a three-peak structure for both, classical and quantum models. The origin of these peaks can be undestood by a detailed analysis of the classical trajectories. In particular for SEF, it was found [23] that the chaotic trajectories dweel for a long time in the neighborhood of stable and unstable period-one orbits. The unstable (stable) period-one orbit is defined by $P_u = 0$ and $x = 0 (\pi)$. The right peak corresponds to motion perpendicular to the $x$ 
axis at $x=\pi $ (unstable orbit). Similarily, the left one results from the stable orbit at $x=0$, see details in Ref. [23]. In contrast, the central peak is produced when the trajectory is parallel to the $x$-axis. A similar analysis explains the origin of the structure of the classical LDOS [23]. One can detect a difference in the structure of the LDOS in comparison to that of the SEF, mainly, in the central region. This fact may be explained by the difference between time averages and phase space averages for finite times (see also discussion in Ref. [25]). It is remarkable that the quantum distributions are sensitive to the existence of these periodic orbits. In general there is an overall good correspondence between classical and quantum distributions (as exemplified by Fig. 4) but with large fluctuations due to the existence of localized and sparse states within the energy shell. 

The kind of localization which we observed in this study occurs in a finite energy range determined by the classical energy shell.  On the other hand, in time-periodic models, localization happens in unbounded momentum or energy space, and chaotic eigenstates are dense in the unperturbed basis. On the contrary, localization in conservative systems reveals itself in two different ways, one is very strong (perturbative) localization, see example in Fig. 2b, and the second one appears in the form of sparse eigenstates, see Fig. 3a. So far, it is not clear whether in conservative systems, the former type of localization can also appear.

	In summary, we have analyzed quantum-classical correspondence for the eigenstates 
and the local spectral density of states for non-integrable billiards, using as a paradigm a 2D channel with periodic sinusoideal boundaries. We have shown that the global shape of the quantum quantities can be determined  from their classical ones. In particular, the range of the energy shell and its global shape can be deduced from the classical distributions. It was also shown that the existence of localized and sparse states within the energy shell give rise to strong quantum fluctuations. These states are also responsible for the observed deviations from random matrix theory predictions for level spacing statistics and are expected to play an important role in the transport properties of the system. We remark that the comparison was done here for quantum and purely classical quantities. Thus, it is very interesting to develop a kind of semiclassical approach for the shape of eigenstates and the LDOS, based on the trace formula (for time-periodic pertubation, this problem has been addressed in Ref. [26]).\\

\newpage
{\Large \bf Acknowledgements}\\

This work was partially supported by  CONACYT (Mexico) grant, No. 26163-E and NSF-CONACYT, grant No. E120.3341.


\begin{thebibliography}{99}
\bibitem{text1} For narrow channels ($d<2 \pi$) global chaos cannot be achieved due to the appearance of a
large resonance island localized in the neighborhood of the widest part of the channel ($x=0$, mod $ 2\pi$). See Ref.[2].
\bibitem{text2}Since the Hamiltonian $H^0$ is not fully chaotic, one needs to perform an additional average over many
different initial conditions belonging to a given energy $E^0$, see [23] for details.
\end{thebibliography}
\end{document}